\documentclass[12pt]{article}
\usepackage{amssymb,amsfonts}
\usepackage[top=2.75cm, bottom=2.75cm, left=2.75cm, right=2.75cm]{geometry}

\newcommand{\be}{\begin{equation}}
\newcommand{\ee}{\end{equation}}
\newcommand{\beqs}{\begin{eqnarray}}
\newcommand{\eeqs}{\end{eqnarray}}
\def\({\left(}
\def\){\right)}

\def\N{ {\cal N}}
\def\Z{ {\cal Z}}

\def\mxth{\mathsurround=0pt }
\def\xversim#1#2{\lower2.pt\vbox{\baselineskip0pt \lineskip-.5pt
x  \ialign{$\mxth#1\hfil##\hfil$\crcr#2\crcr\sim\crcr}}}

\newcommand{\pa}{\partial}

\newcommand{\z}{\zeta}

\newcommand{\pis}{{\pi\kern-1.28ex /}}

\newcommand{\Y}{\Upsilon}
\newcommand{\bY}{{\bar\Upsilon}}

\newcommand{\OO}{ {\cal O}}
\newcommand{\cpl}{$\mathbb{P}^1$ }

\def\+{{+\!\!\!+}} 
\def\pp{\mbox{\tiny${}_{\stackrel\+ =}$}}

\begin{document}
\begin{titlepage}
\begin{flushright} \small
UUITP-04/07 \\ HIP-2007-14/TH.\\ 
\end{flushright}
\smallskip
\begin{center} \LARGE
{\bf Hyperk\"ahler Metrics from Projective Superspace }\normalsize\footnote{Contribution to ``The 27th Winter School GEOMETRY AND PHYSICS'', Czech Republic, Srni, January 13 - 20, 2007}
 \\[12mm] 
{\bf Ulf~Lindstr\"om$^{a,b}$,} \\[8mm]
 {\small\it
$^a$Department of Theoretical Physics 
Uppsala University, \\ Box 803, SE-751 08 Uppsala, Sweden \\
~\\
$^b$HIP-Helsinki Institute of Physics, University of Helsinki,\\
P.O. Box 64 FIN-00014  Suomi-Finland\\
~\\
}
\end{center}
\vspace{10mm}
\centerline{\bfseries Abstract} \bigskip

\noindent  
This is a brief review of how sigma models in Projective Superspace have become important tools
for constructing new hyperk\"ahler metrics.
\vfill
\end{titlepage}
\newpage
\setcounter{page}{1}
\normalsize

\section{Introduction}
\label{intro}
The close relation between supersymmetric sigma models and complex geometry was first observed almost thirty years ago in \cite{Zumino:1979et}.  For $\N=2$ models in four dimensions the target space geometry was subsequently shown to be hyperk\"ahler in \cite{Alvarez-Gaume:1981hm}.  This fact was extensively exploited in a $\N=1$ superspace formualtion of these models in \cite{Lindstrom:1983rt}, where two new constructions were presented; the Legendre transform construction and the hyperk\"ahler quotient construction. The latter reduction was given a mathematical formulation in \cite{Hitchin:1986ea} where we also elaborated on a manifest $\N=2$ formulation, originally introduced in \cite{Gates:1984nk}. 

A $\N=2$ superspace formulation of the $\N=2$ sigma model is obviously desirable, since it will automatically lead to hyperk\"ahler geometry on the target space. The $\N=2$ Projective Superspace
which makes this possible grew out of the development mentioned in the last sentence in the paragraph above. Over the years it has been developed and refined in, e.g., \cite{Karlhede:1984vr}-\cite{Gonzalez-Rey:1997db}. In this article we report on some of that development along with some very recent applications.

\section{Sigma Models}
\label{sigma}
A supersymmetric non-linear sigma model is given by maps from a (super) manifold $\Sigma^{(d,\N)}$ to a target spac ${\cal T}$:
\be
\Phi : \Sigma^{(d,\N)} \longmapsto {\cal{T}}~, 
\ee
defined by giving an action involving an integral over $\Sigma^{(d,\N)}$. For a two-dimensional model in
$\N=(1,1)$ superspace ($d=2, \N=(1,1)$) the action is 
\bigskip
\be
\label{action}
S=\int_\Sigma d^2\xi d^2\theta D_+\Phi^\mu E_{\mu\nu }(\Phi)D_-\Phi^\nu~,
\ee
where  $\xi,\theta$ are coordinates on $\Sigma$, the superspace covariant derivatives satisfy $D^2_\pm=i\pa_{\pp}$, and $E_{\mu\nu }\equiv G_{\mu\nu }+B_{\mu\nu }$ is the sum of the metric and antisymmetric $B$-field. The field equations are
\be
\nabla_+^{(+)}D_-\Phi^\mu=0
\ee
which involves the pullback of the covariant derivative $\nabla^{(+)}\equiv \nabla+G^{-1}H$, the sum of the Levi-Civita connection and the torsion built from the field-strength of the $B$-field; $H=dB$. The r\^{o}le of the geometry of  {$\cal{T}$} is becoming evident from the geometric objects introduced.
The type of geometry depends on $(d,\N)$, i.e., on the bosonic dimension of $\Sigma$ and on the number of supersymmetries. We illustrate with a couple of examples.
\bigskip

{\em Ex.1}\\

The model defined by the action (\ref{action}) has $\N=(2,2)$ supersymmetry provided that the target space carries a certain bi-hermitean geometry \cite{Gates:1984nk}, or in its modern guise, Generalized  K\"ahler Geometry \cite{hitchinCY} \cite{gualtieri}. In this case, there is a manifest $\N=(2,2)$ formulation 
\be
S= \int_M{\mathbb{D}}^2{\mathbb{D}}^2K(\mathbb{X}_L,\bar\mathbb {X}_L, \mathbb{X}_R,\bar\mathbb{X}_R,\phi,\bar\phi,\chi,\bar\chi)~,
\ee
where the Lagrangian $K$ is a function or  the chiral $\phi$ and  twisted chiral fields $\chi$ as well as the semichiral fields \cite{Buscher:1987uw}, $\mathbb{X}_{L,R}$. These fields are defined as follows:
\beqs
\label{constraints}
&&\bar \mathbb{D}_{+}\mathbb{X}_L= \mathbb{D}_+ \bar\mathbb{X}_L = 0~,\cr
&&\cr
&&\bar\mathbb{D}_- \mathbb{X}_R = \mathbb{D}_{-}\bar\mathbb{X}_R = 0~.\cr
&&\cr
&&\bar \mathbb{D}_{\pm}\phi=\mathbb{D} _{\pm}\bar\phi=0\cr
&&\cr
&&\bar\mathbb{D}_{+}\chi= \mathbb{D}_{-}\chi=
\mathbb{D}_{+}\bar\chi=\bar\mathbb{D}_{-}\bar \chi=0~,
\eeqs
where $\mathbb{D}$ is the  ${\cal{N}}=(2,2)$ covariant derivative. All geometric quantities in this geometry have a local expression involving derivatives of the Generalized K\"ahler potential $K$
\cite{Lindstrom:2005zr}. These expressions, in particular those for the metric and $B$-field, are non-linear functions of $\pa\pa K$, nonlinearities that can be explained by the fact that the geometry may be constructed by a quotient from a higher dimensional space \cite{Lindstrom:2007qf}.
\bigskip

{\em Ex.2}\\

Consider the previous example without a $B$-field.
When the number of supersymmetries are further increased to ${\cal{N}}=(4,4)$, the target space geometry is restricted to be hyperk\"ahler. The K\"ahler potential is $K(\phi,\bar \phi)$ and the additional supersymmetries are non-manifest, i.e., explicit transformations of the chiral and semichiral superfields.
These transformations involve the additional two complex structures of the hyperk\"ahler geometry, and the algebra of the extra supersymmetries typically only close on-shell, i.e., modulo field equations.

\section{Projective Superspace}

In the second example above, the ${\cal{N}}=(2,2)$ formulation of the ${\cal{N}}=(4,4)$ models require explicit transformations on the ${\cal{N}}=(2,2)$ superfields that close to the supersymmetry algebra on-shell. This non-manifest formulation makes the construction of new models difficult. Below follows a brief description of a superspace where all supersymmetries are manifest. This ``projective superspace'' \cite{Karlhede:1984vr}-\cite{Gonzalez-Rey:1997db} has been developed in parallel to harmonic superspace 
\cite{Galperin:2001uw}. The relation between the two approaches is discussed in \cite{Kuzenko:1998xm}.

A hyperk\"ahler space ${\cal T}$ supports three globally defined integrable complex structures $I,J,K$ obeying the quaternion algebra: $IJ=-JI=K$, plus cyclic permutations. Any linear combination of these,  $aI+bJ+cK$ is again a complex structure on ${\cal T}$ if $a^2+b^2+c^2=1$, i.e., if $\{a,b,c\}$ lies on a two-sphere $S^2\backsimeq \mathbb{P}^1$.
The Twistor space $\Z$ of  a hyperk\"ahler space ${\cal T}$ is the product of ${\cal T}$ with this two-sphere $\Z= {\cal T}\times\mathbb{P}^1$. 
The two-sphere  thus parametrizes the complex structures and we choose projective coordinates $\zeta$ to describe it (in a patch including the north pole). It is an interesting and remarkable fact that the very same
$S^2$ arises in an extension of superspace to accomodate manifes ${\cal{N}}=(4,4)$ models.

The algebra of ${\cal{N}}=(4,4)$ superspace derivatives is
\beqs
&\{\mathbb{D}_{a\pm},\bar{\mathbb{D}}^b_{\pm}\}=\pm i\delta^b_a\pa_{\pp}~,
&\{\mathbb{D}_{a\pm},\mathbb{D}_{b\pm}\}=0\cr
&\{\mathbb{D}_{a\pm},\mathbb{D}_{b\mp}\}=0~,~~~~~~
&\{\mathbb{D}_{a\pm},\bar\mathbb{D}^b_{\mp}\}=0
\eeqs
We may parameterize a \cpl
of maximal graded abelian subalgebras
as (suppressing the spinor indices)
\be
\label{projder}
\nabla(\z)={\mathbb{D}}_{2}+\z {\mathbb{D}}_{1}~,~~
\bar{\nabla} (\z)
=\bar{\mathbb{D}}^1-\z\bar{\mathbb{D}}^2~,
\ee
where $\z$ is the coordinate introduced above, and the bar on $\nabla$ denotes conjugation with respect to a real structure $\mathfrak{R}$ defined as complex conjugation composed with the antipodal map on
$\mathbb{P}^1\backsimeq S^2$. 
The two new covariant derivatives in (\ref{projder}) anti-commute
\be
\{\nabla ,\bar \nabla\}=0~.
\ee
They may be used to introduce constraints on superfields similarily to how the ${\cal{N}}=(2,2)$ derivatives are used to impose chirality constraints in (\ref{constraints}). Superfields now live in an  extended superspace with coordinates $\xi,\z, \theta$. The superfields $\Y$ we shall be interested in satisfy the projective chirality constraint
\be
\label{projchir}
\nabla \Y=\bar \nabla\Y=0~,
\ee
and are taken to have the folloving $\z$-expansion:
\be
\label{expansion}
\Y=\sum_i\Y_i\z^i~.
\ee
We use the real structure  acting on superfields, $\mathfrak{R}(\Y)\equiv \bar\Y$, to impose reality conditions on the superfields. An $\OO (2n)$ multiplet is thus defined via
\be
\label{otwonmult}
\Y\equiv \eta_{(2n)}=(-)^n\z^{2n}\bar\Y~.
\ee

The expansion (\ref{expansion}) is useful in displaying the ${\cal{N}}=(2,2)$ content of the multiplets.
Using the relation (\ref{projder}) to the ${\cal{N}}=(2,2)$ derivatives in (\ref{projchir}) we read off the following expansion for an $\OO (4)$ multipet (\ref{otwonmult}):
\be
\eta_{(4)}=\phi +\z \Sigma +\z^2X-\z^3\bar\Sigma + \z^4\bar\phi~,
\ee
with the component ${\cal{N}}=(2,2)$ fields being chiral $\phi$, unconstrained $X$ and complex linear $\Sigma$.
A complex linear field satisfies
\be
\bar{\mathbb{D}}^2\Sigma =0~,
\ee
and is dual to a chiral superfield.
A general projective chiral $\Y$ has the expansion
\be
\Y=\phi +\z \Sigma +\sum_{i=2}^\infty X_i\z^i~,
\ee
with all $X_i$'s unconstrained.

\section{The Generalized Legendre Transform}

In this section we review one particular construction of hyperk\"ahler metrics using projective superspace  introduced in \cite{Lindstrom:1987ks}.

An ${\cal{N}}=(4,4)$ invariant action may be written as
\be
\label{act1}
S=\int\mathbb{D}^2\bar{\mathbb{D}}^2F~,
\ee
with 
\be
\label{action2}
F\equiv \oint_C\frac{d\z }{2\pi i \z }f(\Y,\bar\Y;\z)~,
\ee
for some suitably defined contour $C$.  Eliminating the auxiliary fields $X_i$ by their equations of motion will yield an ${\cal{N}}=(2,2)$ model defined on the tangent bundle 
$T(\cal{T})$ parametrized by $(\phi,\Sigma)$. Dualizing the complex linear fields $\Sigma$ to chiral fields $\tilde \phi$ the final result is a supersymmetric ${\cal{N}}=(2,2)$ sigma model in terms of $(\phi, \tilde\phi)$ which is guaranteed by construction to have ${\cal{N}}=(4,4)$ supersymmetry, and thus to define a hyperk\"ahler metric. In equations, these steps are:\\
Solve the equations of motion for the auxiliary fields:
\be\label{dF}
\frac{\pa F}{\pa\Y_i}=
\oint_C\frac{d\z}{2\pi i\z}\,\z^i
\left(\frac\pa{\pa\Y}f(\Y,\bY;\z)\!\right)=0~~,~~~i\ge 2~.
\ee
Solving these equations puts us on ${\cal{N}}=2$-shell, which means that only the ${\cal{N}}=(2,2)$ component symmetry remains off-shell. (In fact, insisting on keeping the ${\cal{N}}=(4,4)$ constraints
(\ref{projchir}) will put us totally on-shell.)
In ${\cal{N}}=(2,2)$ superspace the resulting model, after eliminating $X_i$, is given by a Lagrangian
$K(\phi,\bar\phi,\Sigma,\bar \Sigma)$.  This is dualized to $\tilde{K}(\phi,\bar \phi,\tilde \phi,\bar {\tilde \phi})$ via a Legendre transform
\beqs
&&\tilde{K}(\phi,\bar \phi,\tilde \phi,\bar{\tilde \phi})=K(\phi,\bar \phi,\Sigma,\bar \Sigma)-\tilde \phi \Sigma-\bar{\tilde \phi} \bar \Sigma\cr
&&\cr
&&\tilde \phi=\frac{\pa K}{\pa \Sigma}~,~~~\bar{\tilde \phi}=\frac{\pa K}{\pa \bar \Sigma}~.
\eeqs          

\section{Hyperk\"ahler metrics on Hermitean symmetric spaces}

 This section contains an introduction to the recent paper \cite{Arai:2006gg} where the generalized Legendre transform described in the previous section is used to find metrics on the Hermitean symmetric  spaces listed in the following table:

\bigskip
  \begin{center}
  \begin{tabular}[htb]{|l|l|}
  \hline
  Compact&Non-Compact\cr
  \hline
  $U(n+m)/U(n)\times U(m)$&$U(n,m)/U(n)\times U(m)$\cr
  $SO(2n)/U(n);~Sp(n)/U(n)$&$SO^*(2n)/U(n);~Sp(n,\mathbb{R})/U(n)$\cr
  $SO(n+2)/SO(n)\times SO(2)$& $SO_0(n+2)/SO(n)\times SO(2)$\cr
  \hline
  \end{tabular}
  \end{center}
  \bigskip

The special features of these quotient spaces that allow us to find a hyperk\"ahler metric on their co-tangent bundle is the existence of holomorphic isometries and that we are able to find convenient coset representatives.

A simple example of how the coset representative enters in understanding a quotient is given, e.g., in  \cite{vanNieuwenhuizen:1984ke}: In $\mathbb{R}^{n+1}$
the sphere $S^n$ forms a representation of $SO(n+1)$. The isotropy subgroup at the north pole $p_0$ of $S^n$ is $SO(n)$. Consider another point $p$ on $S^n$ an let $g_p\in SO(n+1)$ be an element that maps $p_0 \to p$. The complete set of elements of  $SO(n+1)$ which map $p_0 \to p$ is thus of the form $g_pSO(n)$, or in other words $S^n=SO(n+1)/SO(n)$. A coset representative is a choice of element in  
 $g_pSO(n)$, and that choice can make the transport of properties defined at the north pole to an arbitrary point more or less transparent.

An important step in the generalized Legendre transform is to solve the auxiliary field equation (\ref{dF}). As outlined in \cite{GK1} and further elaborated in \cite{K3}, 
for Hermitian symmetric spaces 
the auxiliary fields may be eliminated exactly. In the present case, we start from a solution at the origin $\phi=0$,
\be
\label{auxi}
\Y^{(0)}=\z\Sigma^{(0)}~. 
\ee
We then extend this solution to a solution $\Y^*$ at an arbitrary point using a coset representative.
We illustate the method in a simple example due to S. Kuzenko.

{\em Ex. 3}

The K\"ahler potential for \cpl is given by
\be
K(\phi,\bar \phi)=ln(1+\phi\bar \phi)~,
\ee
and we denote the metric that follows from this by $g_{\phi , \bar \phi}$. Here 
$\phi$ 
is a holomorphic coordinate which we extend to an $\N=(2,2)$ chiral superfield.
To construct a  hyperk\"ahler metric we first replace $\phi\to \Y$, and then solve the auxiliary field equation as in (\ref{auxi}). Thinking of 
$\mathbb{C}\mathbb{P} ^n$ as the quotient $G_{1,n+1}({\mathbb C})= {\rm U} (n+1) / {\rm U}(n) \times {\rm U}(1)$, we use a coset representative $L(\phi, \bar \phi)$ to extend the solution from the origin to an arbitrary point. The result is
\be
\Y^*=\frac {\Y^{(0)}+\phi }{1-\Y^{(0)}\bar \phi }=\frac {\z\Sigma^{(0)}+\phi }{1-\z\Sigma^{(0)}\bar \phi }~.
\ee
To find the chiral multiplet $\Sigma$ that parametrizes the tangent bundle, we use the definition
\be
\Sigma \equiv \frac{d\Y^*}{d\z }|_{\z=0}=(1+\phi\bar \phi)\Sigma^{(0)}~,
\ee
yielding
\be
\Y^*=\frac{(1+\phi\bar \phi)\phi+\z\Sigma }{(1+\phi\bar \phi)-\z\Sigma \bar\phi }~.
\ee
The $\N=(2,2)$ superspace Lagrangian on the tangent bundle is then
\be
K(\Y^*,\bar \Y^*)=K(\phi,\bar\phi)+ln(1-g_{\phi\bar\phi}\Sigma\bar \Sigma)~.
\ee   
The final Legendre transform replacing the linear multiplet by a new chiral field, $\Sigma \to \tilde \phi$   produces the K\"ahler potential $K(\phi,\bar \phi, \tilde \phi, \bar {\tilde \phi})$ for the Eguchi-Hanson metric.    
\bigskip

The \cpl example captures the essential id\'{e}a in our construction. The reader is referred to the paper \cite{Arai:2006gg} for details. 

\section{Other alternatives in Projective Superspace} 

Of the two methods for constructing hyperk\"ahler metrics introduced in 
\cite{Lindstrom:1983rt}, we have dwelt on the Legendre transform method and its generalization to projective superspace. The hyperk\"ahler reduction (hyperk\"ahler quotient construction) that we further elaborated on in  \cite{Hitchin:1986ea}, may also be lifted to projective superspace. Both these methods involve only chiral  
$\N=(2,2)$ superfields. When a nonzero $B$-field is present , the $\N=(2,2)$ sigma models involve all the superfields in (\ref{constraints}), as discussed in section \ref{sigma}. For a full description of (generalizations of) hyperk\"ahler metrics on such spaces, the doubly projective superspace \cite{Buscher:1987uw} is required. We now briefly touch on this construction.

In the doubly projective superspace, at each point in ordinary superspace we intrduce one \cpl for each chirality and denote the corresponding coordinates by 
$\z_L$ and $\z_R$. The condition (\ref{projder}) turns into
\beqs
\label{LRprojder}
&&\nabla_+(\z_L)={\mathbb{D}}_{2+}+\z_L {\mathbb{D}}_{1+}~,\cr
&&\cr
&&\nabla_-(\z_R)={\mathbb{D}}_{2+}+\z_R {\mathbb{D}}_{1-}~,
\eeqs
with the conjugated operators defined with respect to the real structure   
$\mathfrak{R}$ acting on both $\z_L$ and $\z_R$. A superfield has the expansion
\be
\label{expansion}
\Y=\sum_{i,j}\Y_{i,j}\z_L^i\z_R^j~,
\ee
and is taken to be both left and right projectively chiral. We may also impose reality conditions using $\mathfrak{R}$, as well as particular conditions on the components, such as the ``cylindrical'' condition
\be
\Y_{i,j+k}=\Y_{i,j}~,
\ee
for some $k$.
Actions are formed in analogy to (\ref{act1}) and (\ref{action2}). The $\N=(2,2)$ components of such a model  include twisted chiral fields $\chi$, as well as semi-chiral
ones $\mathbb{X}_{L,R}$. In fact this is the context in which the semi-chiral $\N=(2,2)$ superfields were introduced \cite{Buscher:1987uw}. Hyperk\"ahler metrics derived in this superspace are discussed in \cite{Lindstrom:1994mw}.
An exciting project is to merge this picture with the recent results in \cite{Lindstrom:2007qf}.
\bigskip

\noindent{\bf\large Acknowledgement}:
\bigskip

I am very happy to acknowledge all my collaborators on the papers that form the basis of this brief report. In particular I am grateful for the many years of collaboration with Martin Ro\v cek and the recent rejuvenating collaborations with Rikard von Unge and Maxim Zabzine.
The work was supported by EU grant (Superstring theory)
MRTN-2004-512194 and VR grant 621-2003-3454.

%%%%%%%%%%%%%%%%%%%%%%%
% ---- Bibliography ----

\end{document}